\documentclass[prb,aps,twocolumn,floatfix,showpacs]{revtex4}
\usepackage{bm}
\usepackage{amssymb}
\usepackage{graphicx}

\def\horparallel{
\begin{picture}(8,7)
\thicklines
\put(1,-0.5){\line(1,0){6}}
\put(1,5.5){\line(1,0){6}}
\end{picture}}

\def\vertparallel{
\begin{picture}(8,7)
\thicklines
\put(1,-0.8){\line(0,1){7}}
\put(6.5,-0.8){\line(0,1){7}}
\end{picture}}

\begin{document}

\title{Effective quantum dimer model for trimerized kagom\'e antiferromagnet}

\author{M. E. Zhitomirsky}

\affiliation{
Commissariat \`a l'Energie Atomique, DSM/DRFMC/SPSMS, 38054 Grenoble,
France}
\date{\today}

\begin{abstract}
An effective spin-orbit Hamiltonian is derived for a 
spin-1/2 trimerized kagom\'e antiferromagnet in the 
second-order of perturbation theory in the ratio of two coupling
constants. Low-energy singlet states of the obtained model
are mapped to a quantum dimer model on a 
triangular lattice. The quantum dimer model is 
dominated by dimer resonances
on a few shortest loops of the triangular lattice.
Characteristic energy scale for the dimer model constitutes only a small
fraction of the weaker exchange coupling constant.
\end{abstract}

\maketitle

\section{Introduction}

Resonating valence bond (RVB) state \cite{anderson} 
is nowadays a popular 
paradigm in condensed matter physics. Short-range RVB states are 
considered to be probable candidates for an elusive spin-liquid 
phase of magnetic insulators. The idea of short-range RVB states is 
quantitatively formulated by so called quantum dimer models (QDM).
\cite{sutherland,rokhsar,moessner00}
In a QDM each dimer represents a singlet state (valence bond) 
between a pair of neighboring spins.
The QD Hamiltonian is defined in the Hilbert space
of close-packed dimer coverings of a lattice. The dimer states
are assumed
to be properly orthogonalized. Local dynamics of an RVB state
is typically described on the smallest 
plaquettes ($\square$), which are squares for a square lattice
or rhombi for a triangular lattice:
\begin{equation}
\hat{\cal H}_{\rm QD}=\sum_\square \Bigr[-t \bigl(
|\horparallel\rangle\langle \vertparallel|+
|\vertparallel\rangle\langle\horparallel|\bigr)
+ V \bigl(
\left| \horparallel  \right\rangle\left\langle \horparallel  \right|+
\left| \vertparallel \right\rangle\left\langle \vertparallel \right| \bigr)
\Bigr].
\label{Hqdm}
\end{equation}
The first term is a dimer kinetic energy, which flips
a pair of parallel dimers around an arbitrary plaquette; 
the second term is a potential
energy between such pairs. 

Rokhsar and Kivelson \cite{rokhsar} have shown that a short-range 
RVB state given 
by a superposition of all dimer coverings of 
a square lattice is an exact eigenstate of the QD Hamiltonian 
for a special choice of the parameters $t=V$.
On a bipartite square lattice, the RVB state 
at the Rokhsar-Kivelson (RK) point has long-range power low correlations
and describes, consequently, a gapless spin-liquid state.
\cite{rokhsar,moessner01} Small 
perturbations away from the RK point drive the system into one 
of the ordered crystalline dimer states.
The QDM on a triangular lattice exhibits quite a different 
behavior at the RK point. \cite{moessner01,ioselevich}
The short-range RVB state has exponentially decaying correlators and
is fully gapped. It exists, therefore, in a finite range 
of parameters around the RK point
and is stable with respect to weak perturbations
to the QD Hamiltonian.
Still, question whether such states or Hamiltonians can 
describe realistic quantum spin systems  
remains unsettled.
In the present work we propose 
a realization of QDM on a triangular lattice for a nearest-neighbor
Heisenberg spin model.

The most probable candidates for a singlet spin-liquid ground state 
are frustrated quantum antiferromagnets. \cite{book} 
Numerical exact diagonalization studies of a spin-1/2 Heisenberg
kagom\'e antiferromagnet have shown that this spin model has 
a nonmagnetic ground state with a large number of low-lying
singlet excitations. \cite{lecheminant,waldtmann} 
Accessible cluster sizes do not allow to draw 
a definite conclusion on the possible nature of the magnetically disordered
(singlet) ground state.

\begin{figure}
\centerline{
\includegraphics[width=0.9\columnwidth]{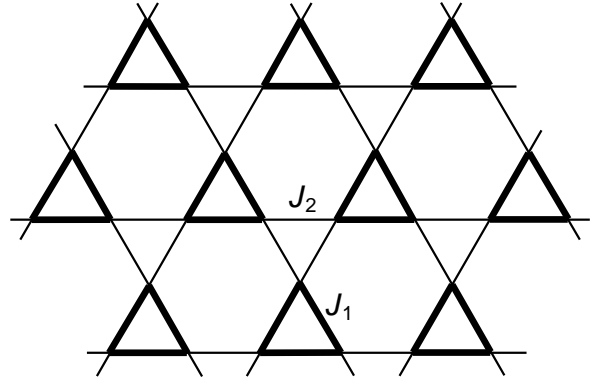}}
\caption{Trimerized  kagom\'e lattice with two exchange
constants: $J_1$ in $\bigtriangleup$-triangles and
$J_2$ in $\bigtriangledown$-triangles.}
\label{trimerized}
\end{figure}

One of a very few analytic approaches to such strongly-correlated
spin systems is an 
expansion from small clusters. The main motif 
of a kagom\'e lattice is triangle. It is, therefore, natural to 
start from a trimerized kagom\'e lattice shown in Fig.~\ref{trimerized}.
Such a strong-coupling approach has been pursued in relation
to kagom\'e antiferromagnet in several theoretical works.
\cite{subrah,mila,raghu,auerbach}
Recently, an experimental scheme to create a trimerized kagom\'e lattice
has been suggested for ultracold atomic gases in optical traps.
\cite{santos} This opens a way for an experimental
probe of RVB physics in the corresponding spin model.

The Heisenberg model on a trimerized lattice
\begin{equation}
\hat{\cal H} = \sum_{\langle ij\rangle} J_{ij} {\bf S}_i\cdot{\bf S}_j
\label{Hheis}
\end{equation}
has two coupling constants: 
$J_1$ for a stronger interaction between spins in 
$\bigtriangleup$-triangles and $J_2$ for a weaker interaction 
inside $\bigtriangledown$-triangles.
An array of isolated $\bigtriangleup$-blocks is a zeroth order 
Hamiltonian, which has a highly degenerate ground state.
Interblock interaction lifts
such a degeneracy. In section II we derive an effective 
Hamiltonian up to the second-order in a small parameter $J=J_2/J_1\ll 1$. 
This Hamiltonian is mapped to a QDM in section III.
The obtained results and their implication for the ground state
properties of the trimerized kagom\'e model are discussed in 
section IV.

\section{Strong-coupling expansion}

\subsection{First-order Hamiltonian}

\begin{figure}
\centerline{
\includegraphics[width=0.9\columnwidth]{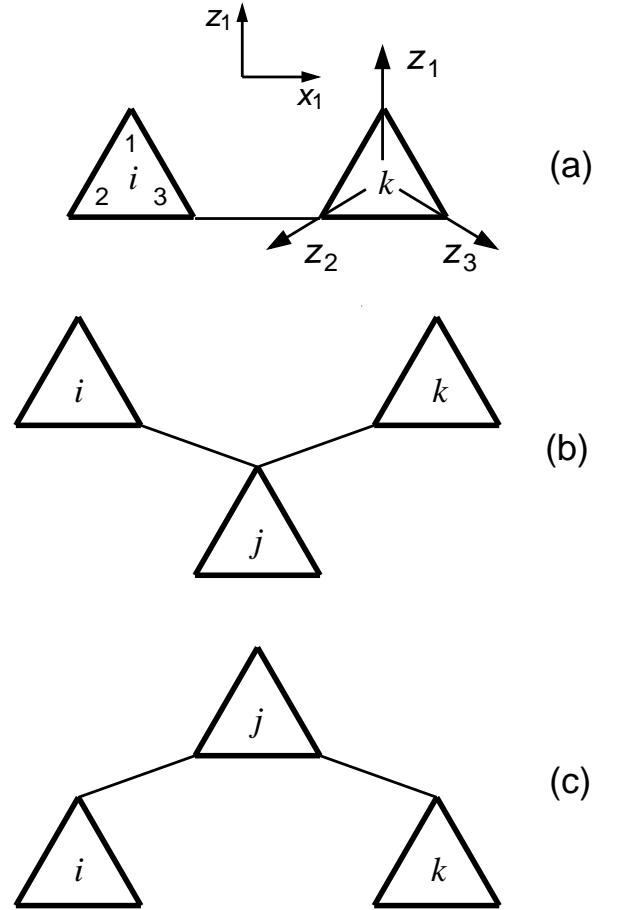}}
\caption{Three different geometries of $\bigtriangleup$-blocks
contributing to the second-order energy correction
in the interblock coupling.
The labeling of axes and sites inside triangles is shown in 
the upper panel (a). }
\label{strong}
\end{figure}

Let us in the beginning rederive the previous results 
on the effective first-order Hamiltonian \cite{subrah,mila,raghu}
using somewhat different notations. Below 
we normalize all energies to $J_1$ such that $J_2\rightarrow J$.
In a strong-coupling
expansion one 
starts with an isolated triangle described by
\begin{eqnarray}
\hat{\cal H}_\bigtriangleup & = & {\bf S}_1\cdot{\bf S}_2 
+ {\bf S}_2\cdot{\bf S}_3 + 
{\bf S}_3\cdot{\bf S}_1  \nonumber \\
& = &  \frac{1}{2}\bigl({\bf S}_1 +{\bf S}_2+{\bf S}_3\bigr)^2
- \frac{3}{2} S(S+1) \ .
\end{eqnarray} 
The energy levels of $\hat{\cal H}_\bigtriangleup$ are determined by 
the total spin $S_{\rm tot}$. For on-site $S=1/2$,
which is always assumed below, the levels are 
two degenerate doublets with $S_{\rm tot}=1/2$ and $E=-\frac{3}{4}$ and 
one quartet with $S_{\rm tot}=3/2$ and $E=\frac{3}{4}$.
The doublet states with $S^z_{\rm tot}=1/2$ are
\begin{equation}
|d_\uparrow\rangle= 
\frac{1}{\sqrt{2}}\bigl(\uparrow\uparrow\downarrow\!-\!
\uparrow\downarrow\uparrow \bigr),\ \
|p_\uparrow\rangle=\frac{1}{\sqrt{6}}\bigl(2\!\downarrow\uparrow\uparrow 
\!-\!\uparrow\uparrow\downarrow\!-\!
\uparrow\downarrow\uparrow \bigr),
\label{doublet}
\end{equation}
where spin numbering in an individual triangle
follows Fig.~\ref{strong}a. The former state  
$|d_\uparrow\rangle$ is a combination of the spin-up apex spin and 
a singlet bond between the two base spins and has odd parity under
the permutation $\hat{P}_{23}$. The other state $|p_\uparrow\rangle$
is even under  $\hat{P}_{23}$. 
The two members of a quartet with $S_{\rm tot}^z=+3$ 
and $+1$ are
\begin{equation}
|q_{+3}\rangle= 
|\!\uparrow\uparrow\uparrow\rangle \ , \ \ \ 
|q_{+1}\rangle=\frac{1}{\sqrt{3}}\bigl(
\downarrow\uparrow\uparrow + 
\uparrow\downarrow\uparrow +
\uparrow\uparrow\downarrow \bigr)\ .
\label{quartet}
\end{equation}
All other states are obtained from (\ref{doublet}) and (\ref{quartet})
by applying
$S^-_{\rm tot}$ operator. 

The choice of the basis in the doublet subspace is not unique.
The apex spin can be put into 
a singlet state either with its right or left 
neighbor. The two alternative basis states obtained 
by rotating $|d_\uparrow\rangle$
about a center of triangle counterclockwise
on $2\pi/3$ and $4\pi/3$ are 
\begin{equation}
|d'_\uparrow\rangle=
\frac{1}{\sqrt{2}}\bigl(\downarrow\uparrow\uparrow-\uparrow\uparrow\downarrow
\bigr) \ , \ \ 
|d''_\uparrow\rangle=
\frac{1}{\sqrt{2}}\bigl(\uparrow\downarrow\uparrow-\downarrow\uparrow\uparrow\bigr)
\end{equation} 
with their orthogonal partners 
$|p'_\uparrow\rangle$ and $|p''_\uparrow\rangle$.
Transformation from the old 
basis (\ref{doublet}) to the new states is
\begin{eqnarray}
|d_{\alpha}'\rangle = 
 -\frac{1}{2}|d_{\alpha}\rangle+\frac{\sqrt{3}}{2}|p_{\alpha}\rangle,\   
 |p_{\alpha}'\rangle =
 -\frac{\sqrt{3}}{2}|d_{\alpha}\rangle-\frac{1}{2}|p_{\alpha}\rangle, &&
 \nonumber \\
|d_{\alpha}''\rangle =
-\frac{1}{2}|d_{\alpha}\rangle-\frac{\sqrt{3}}{2}|p_{\alpha}\rangle,\ \ \
 |p_{\alpha}''\rangle = 
\frac{\sqrt{3}}{2}|d_{\alpha}\rangle-\frac{1}{2}|p_{\alpha}\rangle, && 
\label{doublet2}
\end{eqnarray}
where $\alpha=\uparrow,\downarrow$ or 1,2  is a spinor index.
The main difference with the previous works \cite{subrah,mila}
is that real basis states (\ref{doublet}) or (\ref{doublet2})
are used instead of complex chiral states. This yields a more
transparent form of the effective Hamiltonian and simplifies
subsequent derivation of a QDM.

At this point we introduce two sets of the Pauli matrices: 
$\sigma^i$, which act between spin-up and 
spin-down states, and $\tau^i$, which act in the orbital subspace 
$(d,p)$ preserving the total spin.  A convenient
choice of orbital axes shown in Fig.~\ref{strong}a 
corresponds to
\begin{equation}
\tau^{z_1} |d_{\alpha}\rangle = |d_{\alpha}\rangle \ ,\ \ \ 
\tau^{z_1} |p_{\alpha}\rangle =-|p_{\alpha}\rangle \ .
\end{equation}
Then, the orbital operators projected onto the rotated axes 
\begin{equation}
\tau^{z_2}=-\frac{1}{2}\tau^{z_1}
-\frac{\sqrt{3}}{2}\tau^{x_1}, \ \ 
\tau^{z_3} = -\frac{1}{2}\tau^{z_1}+\frac{\sqrt{3}}{2}\tau^{x_1}
\end{equation}
have the following eigenstates:
\begin{eqnarray}
&& \tau^{z_2}|d_{\alpha}'\rangle=|d_{\alpha}'\rangle \ , \ \ \ 
\tau^{z_2}|p_{\alpha}'\rangle=-|p_{\alpha}'\rangle \ ,  
 \nonumber \\
&& \tau^{z_3}|d_{\alpha}''\rangle=|d_{\alpha}''\rangle \ , \ \ \ 
\tau^{z_2}|p_{\alpha}''\rangle=-|p_{\alpha}''\rangle \ .  
\end{eqnarray}

In order to find the effect of interblock coupling 
in the first order of perturbation theory in $J=J_2/J_1$
one should neglect $S_{\rm tot}=3/2$ states separated by a finite gap
$\Delta E = \frac{3}{2}$  and calculate matrix elements
of the on-site spin operators between the low-energy doublet states 
$|d_{\alpha}\rangle$ and $|p_{\alpha}\rangle$. This problem is greatly
simplified once all symmetries are taken into account. 
Introducing operators $d^\dagger_\alpha |0\rangle =
|d_\alpha\rangle$ and  $p^\dagger_\alpha |0\rangle=|p_\alpha\rangle$,
where $|0\rangle$ is a fictitious vacuum, the Hubbard-type representation
of on-site spins is written as 
\begin{eqnarray}
&& {\bf S}_1 = \frac{1}{2}\,d^\dagger_{\alpha}
\mbox{\boldmath$\sigma$}_{\alpha\beta}d_{\beta} -
\frac{1}{6}\,p^\dagger_{\alpha}\mbox{\boldmath$\sigma$}_{\alpha\beta}p_{\beta} \ ,  
\label{rep_spinor} \\
&&
{\bf S}_{2,3} = \frac{1}{3}\,p^\dagger_{\alpha}
\mbox{\boldmath$\sigma$}_{\alpha\beta} p_{\beta}
\pm\frac{1}{2\sqrt{3}}
\left(p^\dagger_{\alpha}\mbox{\boldmath$\sigma$}_{\alpha\beta}d_{\beta}+{\rm h.\,c.}\right).
\nonumber 
\end{eqnarray}
Spinor structure is a consequence of the spin-rotation symmetry, 
while permutation of the base spins
$\hat{P}_{23}$ fixes the orbital part in (\ref{rep_spinor}).
The above representation is further simplified once the total spin 
of a triangle ${\bf S}=\frac{1}{2}\mbox{\boldmath $\sigma$}$ is defined 
and the orbital operators $\tau^{z_k}$ are used. Then,
the $n^{\rm th}$ spin ($n=1$--$3$) of the $i^{\rm th}$ 
triangular block is represented by
\begin{equation}
{\bf S}_{ni} = \frac{1}{3}\,{\bf S}_i\left( 1+2\tau^{z_n}_i\right),
\label{rep_orbit}
\end{equation}
where $\hat{\bf z}_n$ goes from the center of a triangle in 
the direction of the corresponding spin, see Fig.~\ref{strong}a.

The effective first-order Hamiltonian in the interblock coupling 
is found by substituting Eq.~(\ref{rep_orbit})
into the Hamiltonian (\ref{Hheis}):
\begin{equation}
\hat{\cal H}_1 = \frac{J}{9} \sum_{\langle ij\rangle} 
{\bf S}_i\cdot{\bf S}_j(1+2\tau_i^{z_{n}})
(1+2\tau_j^{z_{m}}) \ , 
\label{H1st}
\end{equation}
where a trivial constant term $-\frac{3}{4}N_\bigtriangleup$
is omitted for convenience.
The derived spin-orbital Hamiltonian $\hat{\cal H}_1$ is defined on 
a triangular lattice, such that every site corresponds to
one $\bigtriangleup$-block of the trimerized kagom\'e model
and is attributed with spin and orbital operators.
The bond orientation uniquely determines the orbital
axes for two participating sites.

\subsection{Second-order Hamiltonian}

The second-order energy corrections 
for weakly-coupled spin triangles have been obtained by 
Raghu and co-workers.\cite{raghu} These authors have mostly been 
interested in a one-dimensional model, therefore, their analysis misses 
several terms relevant for a two-dimensional array of triangles
in the trimerized kagom\'e model. In order to calculate 
the second-order result in $J_2$ one needs to determine matrix 
elements of on-site spins between doublet (\ref{doublet})
and quartet (\ref{quartet}) states.
Introducing symmetric third-rank spinor tensor $q_{\alpha\beta\gamma}$
such that $q_{111}=|q_{+3}\rangle$, $q_{112}=\frac{1}{\sqrt{3}}|q_{+1}\rangle$ 
$q_{122}=\frac{1}{\sqrt{3}}|q_{-1}\rangle$, and $q_{222}=|q_{-3}\rangle$,
and utilizing spin-rotation 
symmetry we find by analogy with Eq.~(\ref{rep_spinor})
\begin{eqnarray}
&& {\bf S}_1 = -\frac{i}{\sqrt{6}}\,q^\dagger_{\alpha\beta\gamma}
\bigl(\mbox{\boldmath$\sigma$}\sigma^y\bigr)_{\beta\gamma}p_{\alpha} 
+ {\rm h.\,c.}\ ,
\label{rep_spinor2} \\
&&
{\bf S}_{2,3} = \frac{i}{\sqrt{6}}\,q^\dagger_{\alpha\beta\gamma}
\bigl(\mbox{\boldmath$\sigma$}\sigma^y\bigr)_{\beta\gamma}
\Bigl(\pm\frac{\sqrt{3}}{2}d_\alpha + \frac{1}{2}p_{\alpha}\Bigr) 
+ {\rm h.\,c.}
\nonumber 
\end{eqnarray}

The second-order energy correction in the interblock 
coupling is, generally, given by
\begin{equation}
\hat{\cal H}_2(G,G') = \sum_X \frac{\langle G|\hat{\cal H}|X\rangle
\langle X|\hat{\cal H}|G'\rangle}{E_G-E_X} \ ,
\label{2nd}
\end{equation}
where $G$ and $G'$ denote combinations of lowest doublet
states on $\bigtriangleup$-blocks and $X$ are excited states.
The nonzero second-order terms appear if either
(i) one $J_2$-bond acts twice in the numerator of Eq.~(\ref{2nd})
or (ii) two adjacent $J_2$-bond emerging from the same
$\bigtriangleup$-block are used subsequently in the 
matrix elements $\langle G|\hat{\cal H}|X\rangle$
and $\langle X|\hat{\cal H}|G'\rangle$. This determines
three different geometries for two- and three-block 
interaction terms shown in Fig.~\ref{strong}.
In the first case of two-block interaction, Fig.~\ref{strong}a,
either one or both
triangular blocks have quartets in the intermediate states $X$.
For the three block interactions (Fig.~\ref{strong}b,c),
only a middle block has excited quartets in the intermediate states.
Every pair of free (uncoupled) spins in a $\bigtriangleup$-block imposes
an extra permutation symmetry on the second-order Hamiltonian
$\hat{\cal H}_2(G,G')$.
For example, the two-block cluster in Fig.~\ref{strong}a has extra
$\hat{P}_{12}$ symmetry for the left $i$-block 
and $\hat{P}_{13}$ symmetry for the right $k$-block. 
Therefore, the orbital states, $|d''_\alpha\rangle$
or $|p''_\alpha\rangle$ for the left block and 
$|d'_\alpha\rangle$
or $|p'_\alpha\rangle$ for the right block, remain unchanged
during the second-order perturbation process (\ref{2nd}).
In other words $\hat{\cal H}_{2a}$ commutes with $\tau_i^{z_3}$
and $\tau_k^{z_2}$.
The conservation of orbital state is also fulfilled for all 
triangles in the three-block term in Fig.~\ref{strong}b
and for the side triangles in Fig.~\ref{strong}c, whereas the middle
block does change its orbital state during the second-order
process. The above conservation laws significantly simplify summation
over intermediate states in Eq.~(\ref{2nd}).  The final results are
\begin{eqnarray}
\hat{\cal H}_{2a} & = & -\frac{J^2}{54} \sum_{\langle ik\rangle} 
\Bigl[(3+4\,{\bf S}_i\cdot{\bf S}_k)(1-\tau_i^{z_l}\tau_k^{z_m}) 
\nonumber \\ & & \mbox{}\ \ \ \ \ \ \ \ \ \ +
(1-\tau_i^{z_{n}})(1-\tau_k^{z_{m}})\Bigr] 
\label{H2a}
\end{eqnarray}
for the two-block interaction;
\begin{equation}
\hat{\cal H}_{2b}= -\frac{4J^2}{243}\!\sum_{\langle ijk\rangle} 
{\bf S}_i\cdot{\bf S}_k(1+2\tau_i^{z_l})(1+2\tau_k^{z_m})
(1-\tau_j^{z_n})
\label{H2b}
\end{equation}
for the three-block interaction shown in Fig.~\ref{strong}b
with corresponding labeling of blocks;
and
\begin{eqnarray}
\hat{\cal H}_{2c} & = & \frac{2J^2}{243}\sum_{\langle ijk\rangle} 
(1+2\tau_i^{z_n})(1+2\tau_k^{z_n})\Bigl[{\bf S}_i\cdot{\bf S}_k
(1+2\tau_j^{z_n})
\nonumber \\ & & \mbox{}\ \ \ \ \ \ \ \ \ \ \ + \sqrt{3}\,\tau_j^y\:
{\bf S}_i\cdot({\bf S}_j\times{\bf S}_k)\Bigr]
\label{H2c}
\end{eqnarray}
for the three-block interaction in geometry of Fig.~\ref{strong}c.
Polarization of orbital operators $\tau_i^{z_n}$ in 
the above equations is again found by simple inspection 
of the arrangement of corresponding blocks
on a trimerized kagom\'e lattice. Three-block terms in 
$\hat{\cal H}_{2b}$ exist on $\bigtriangledown$-plaquettes
of an effective triangular lattice with three different
$z$-axes in Eq.~(\ref{H2b}). Three-body terms in $\hat{\cal H}_{2c}$
appear on $\bigtriangleup$-plaquettes of the triangular
lattice with one polarization of $\tau^z$ operators
in Eq.~(\ref{H2c}), which changes under permutation of $(ijk)$.

The interactions $\hat{\cal H}_{2a}$ and $\hat{\cal H}_{2c}$ coincide
up to a trivial change of notations with the previously derived terms 
for a one-dimensional array of triangles,\cite{raghu} while the term 
$\hat{\cal H}_{2b}$ is a novel one.
Note, that $\hat{\cal H}_{2c}$ contains a remarkable three-body spin-chiral
interaction term. By deriving the effective spin-orbital Hamiltonians
$\hat{\cal H}_1$ and $\hat{\cal H}_2$,
we have substantially restricted the Hilbert space and 
simplified the original problem of finding the ground and the lowest
energy states of the spin model (\ref{Hheis}). The remaining problem
of solving $\hat{\cal H}_1+\hat{\cal H}_2$ 
is still highly nontrivial. 
The spin-orbital Hamiltonian $\hat{\cal H}_1$ has been studied so far in
the mean-field approach. \cite{mila}
An effective Hamiltonian derived by the contractor renormalization group,
which partially resembles $\hat{\cal H}_1+\hat{\cal H}_2$,
has also been analyzed in the mean-field approximation. \cite{auerbach}
Below we discuss a mapping of the derived Hamiltonians (\ref{H1st})
and (\ref{H2a})--(\ref{H2c}) to an effective QDM.
The obtained QD Hamiltonian is dominated by the kinetic
energy for dimer tunneling. The mean-field approximation, which assumes 
a frozen pattern
of dimers, is, therefore, a poor approximation in the present problem.

\section{Quantum dimer model}

\subsection{General remarks}

Search for the low-energy states of the first-order Hamiltonian (\ref{H1st}) 
can be started by considering first a two-site problem  (two adjacent
$\bigtriangleup$-blocks of the original kagom\'e lattice). \cite{mila} 
This problem is solved exactly and its ground state 
corresponds to a spin singlet with 
the orbital degrees fully polarized along the bond: 
$\langle \tau_i^{z_n}\rangle =\langle \tau_j^{z_m}\rangle =1$.
The ground-state energy is $-\frac{3}{4}J$.
A variational singlet ground state for the lattice problem (\ref{H1st}) 
is constructed by splitting the whole lattice into a close-packed
structure of dimers between nearest-neighbor sites, such that the dimer
wave-function is given by the ground-state solution of the two-site problem. 
A remarkable feature of these variational states is that
at the mean-field level with respect to orbital degrees
of freedom the total energy is just a sum of energies of individual
dimers and does not depend on a particular dimer covering of
a triangular lattice. \cite{mila}
Indeed, once $\langle\tau_i^{z_n}\rangle=1$,
then for the two other axes $\langle\tau_i^{z_m}\rangle
=\langle \tau_i^{z_k}\rangle \equiv -\frac{1}{2}$. 
Therefore, the expectation value of any empty
bond, {\it i.e.\/}, a bond without a dimer, identically vanishes
over the variational wave-function: either one or both sites 
of the bond have $\langle 1+2\tau_i^{z_m}\rangle\equiv 0$.

The degenerate set of variational mean-field states has 
been identified with low energy states of spin-1/2 antiferromagnets 
on trimerized and isotropic kagom\'e lattices. 
\cite{mila,mambrini} The number of low-lying singlets of 
the kagom\'e model scales, then, as $1.15^N$ in good agreement
with the full exact diagonalization study. \cite{waldtmann}
The previous works leave, however, without answer 
question about validity of the mean-field approximation
and further lifting of degeneracy by quantum fluctuations.

\begin{figure}
\centerline{
\includegraphics[width=0.9\columnwidth]{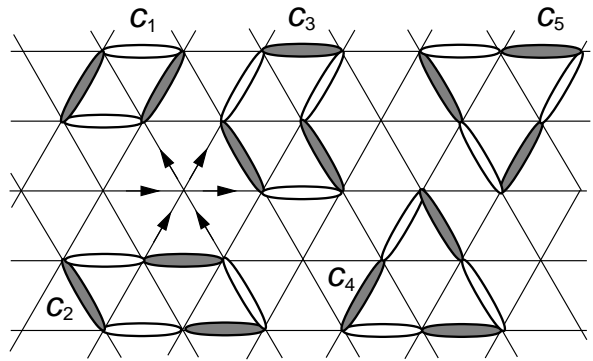}}
\caption{Effective triangular lattice with five 
shortest loops. The arrow directions indicate the sign convention
for singlet wave-functions on each bond}
\label{triangular}
\end{figure}

In order to beyond the mean-field approximation, one has to consider 
off-diagonal matrix elements of the Hamiltonian (\ref{H1st}) between 
various dimer configurations as well as the corresponding overlap
matrix. The general rule to compute the overlap matrix for models, 
where every dimer represents a singlet pair, is to construct 
transition or overlap graph by drawing two dimer configurations on the
same lattice. \cite{sutherland} Every closed nonintersecting 
loop of dimers contributes $2/2^{l/2}$ 
to the overlap matrix, $l$ being the length of the loop. 
The sign of the overlap matrix element depends on a sign convention
for singlet wave-functions. We adopt the standard convention 
\cite{sutherland,moessner01,moessner02} 
such that the singlet bond wave-function is 
$[ab]=\frac{1}{\sqrt{2}}(\uparrow_a\downarrow_b-\downarrow_a\uparrow_b)$,
where $b$ is an upper site in the pair
or is directly to the right from $a$, see Fig.~\ref{triangular}.

Local dynamics of singlet bonds 
in trimerized kagom\'e model is determined by
a few shortest loops on an 
effective triangular lattice, which include two- and three-dimer
moves, see Fig.~\ref{triangular}. 
Taking into account the orbital part of the wave-functions
the overlaps of two dimer configurations on each loop are
calculated as $c_1=-1/2^4$, $c_2=c_3=-1/2^7$, $c_4=-1/2^8$, and 
$c_5=1/2^5$.
These overlap matrix elements are significantly smaller than for singlet 
bond configurations on the original triangular lattice. In the latter case 
the corresponding loops have $c_1=1/2$,
$c_2=c_3=1/2^2$, $c_4=c_5=-1/2^2$.
The difference reflects the fact that loops on an effective
triangular lattice correspond to significantly longer loops
on the original trimerized kagom\'e lattice.
For example, the shortest $C_1$ loop corresponds to a loop
of length $l=10$ 
on a kagom\'e lattice.
Loops $C_4$ and $C_5$ are different for the
considered model because kagom\'e lattice has only
a three-fold rotation axis in the center of every triangle.
Significant difference of the overlap matrix elements explains  
why a QDM description is a poor 
approximation for a spin-1/2 Heisenberg antiferromagnet on a triangular 
lattice,\cite{anderson} but may be a good one for the 
trimerized kagom\'e antiferromagnet.
Below in this section we assume that the ground states of 
the first and second-order effective Hamiltonians 
are given variationally by close-packed dimer states
and compute a new effective QD Hamiltonian.
The above assumption is supported by numerical
treatment of the trimerized kagom\'e antiferromagnet.\cite{mambrini}

Derivation of a QDM from a particular spin Hamiltonian 
has been formulated via calculation of the 
inverse square root of the overlap matrix. \cite{rokhsar} 
We find that actual calculations become more transparent
by operating with the wave-functions. The final result
are, of course, equivalent in both approaches.
Specifically, let us consider two linearly independent, normalized 
states $|\psi_1\rangle$ and $|\psi_2\rangle$, which have a small 
overlap $\langle\psi_1|\psi_2\rangle=\langle\psi_2|\psi_1\rangle=c$. 
Matrix elements of the Hamiltonian between the
two states are assumed to be known 
$E_{11}=\langle\psi_1|\hat{H}|\psi_1\rangle$,
$E_{21}=E_{12}=\langle\psi_1|\hat{H}|\psi_2\rangle$, 
and $E_{22}=\langle\psi_2|\hat{H}|\psi_2\rangle$. The aim is to 
compute matrix elements in a new properly 
orthogonalized basis $|\varphi_n\rangle$. Transformation to the new
basis is given by a symmetric matrix:
\begin{equation}
|\varphi_1\rangle = \lambda\bigl(|\psi_1\rangle - 
\mu|\psi_2\rangle\bigr), \  
|\varphi_2\rangle = \lambda\bigl(|\psi_2\rangle - 
\mu|\psi_1\rangle\bigr)\ . 
\label{phi2}
\end{equation}
Conditions $\langle\varphi_1|\varphi_2\rangle=0$, 
$\langle\varphi_1|\varphi_1\rangle=
\langle\varphi_2|\varphi_2\rangle=1$ determine $\mu$ and $\lambda$. 
Assuming $E_{11}=E_{22}$, for simplicity, and calculating matrix 
elements of $\hat{\cal H}$ between the new states one finds 
\begin{eqnarray}
&&\tilde{E}_{11} = E_{11}+\frac{c}{1-c^2}\bigl( cE_{11}-E_{12}\bigr)\ , 
\nonumber \\
&&
\tilde E_{12}=E_{12}+\frac{c}{1-c^2}\bigl( cE_{12}-E_{11}\bigr)\ .
\label{Eortho}
\end{eqnarray}

In the following we shall subtract from the effective Hamiltonians
$\hat{\cal H}_1$ and $\hat{\cal H}_1+\hat{\cal H}_2$ the corresponding mean-field
energies $E_{\rm MF}$, which are the same for all dimer coverings.
Then, $E_{11}=E_{22}=0$ and the matrix elements (\ref{Eortho}) 
are directly related to the parameters of a QDM:
\begin{equation}
t = -\tilde{E}_{12} \approx - E_{12}\ , \ \ \
V = \tilde{E}_{11} \approx -c E_{12} \ ,
\label{Vt2}
\end{equation}
where $|c|\ll 1$ is used.

\subsection{First-order mapping}

Let us apply the outlined procedure to the first-order Hamiltonian
(\ref{H1st}). The mean-field energy $E_{\rm MF}=-\frac{3}{4}JN_d$ of 
an arbitrary configuration of $N_d$ dimers 
is always subtracted from Eq.~(\ref{H1st}).
The wave-functions for two dimer states 
on the shortest loop $C_1$ shown in Fig.~\ref{triangular}
are written explicitly as
\begin{equation}
|\psi_1\rangle=[12][43]|d_1d'_2d'_3d_4\rangle, \ 
|\psi_2\rangle=[32][41]|d'_1d'_2d''_3d''_4\rangle, 
\end{equation}
where sites are numbered counter-clockwise beginning with the 
lower right corner of the rhombus.
The first part of $|\psi_{1,2}\rangle$ is given by a product
of two spin singlet states, whereas the second part is 
an orbital wave-function represented as a product of states 
(\ref{doublet}) or (\ref{doublet2}).
Explicit calculation of the off-diagonal matrix element
$E_{12} = \langle\psi_2|\hat{\cal H}_1|\psi_1\rangle$ 
yields
\begin{equation}
t_1 = - \frac{3}{2^6} J \ , \ \ \ \ \
V_1 =\frac{3}{2^{10}}J \ .
\label{c1}
\end{equation}
The largest kinetic energy term in the QD Hamiltonian (\ref{Hqdm})
amounts to less
than 5\% of the weaker coupling constant.
The ratio of 
the potential energy to the kinetic term constant is also very small 
$V/|t|=1/16$. 
Note, that $t<0$ from the above calculation. 
Remaining freedom in the choice of sign of tunneling matrix elements
is discussed in the next section.

Since the potential energy $V_1$ is an order of magnitude
smaller than the dimer hopping $t_1$, 
the next relevant interactions besides the kinetic energy
of two-dimer moves around $C_1$ 
may be three-dimer resonances along longer loops $C_2$, $C_3$, $C_4$,
and $C_5$. 
We find no tunneling for loops 
$C_4$ and $C_5$, {\it i.e.},  $V_{4,5}=t_{4,5}\equiv 0$.
For loop $C_5$, vanishing of the off-diagonal matrix element 
can be understood
by drawing two dimer configurations on a corresponding 
cluster of a kagom\'e lattice, which is a six-point star.
Two dimer states around perimeter of such 
a star are exact degenerate eigenstates \cite{syromyat}
and, hence, $t_5\equiv 0$.
In the former case, loop $C_4$, the  
$E_{12}=0$ result is valid only in the first order in $J$, see the 
next subsection.

Coherent motion of three dimers along composite loops $C_2$ and $C_3$
is not described by Eqs.~(\ref{Eortho}) and (\ref{Vt2}) because of
resonances around small rhombi.
Let the additional state with three parallel dimers on the parallelogram
$C_2$ be denoted by $|\psi_1\rangle$ and 
the two dimer states on the perimeter
of $C_2$ be
$|\psi_2\rangle$ and $|\psi_3\rangle$,  $\langle\psi_{2,3}|\psi_1\rangle=c$, 
while $\langle\psi_2|\psi_1\rangle=c'$. 
Then, the matrix elements $E_{12}=E_{13}$ describe short-loop resonances,
while $E_{23}$ corresponds to a tunneling along the composite loop.
Generalizing transformation (\ref{phi2}) to three
states we finally obtain:
\begin{eqnarray}
&&\tilde{E}_{11} \approx - c E_{12} - \bigl(c'-\frac{3}{4}c^2\bigr)E_{23}
\ , 
\nonumber \\
&&
\tilde E_{12} \approx E_{12}  \ , \ \  
\tilde E_{23} \approx  E_{23}- c E_{12} 
\label{Eortho3}
\end{eqnarray}
in the relevant case $|c'|\ll|c|\ll 1$.
Tunneling of dimers along a longer loop $E_{23}$ is renormalized
by short-loop hopping.
Using the above expressions to calculate resonance 
of singlet bonds on $C_2$ and $C_3$ one obtains that in both cases
\begin{equation}
t_2 = t_3 = - \tilde{E}_{23} = - \frac{15}{2^{10}} J \ .
\end{equation}
The potential energy given
by the second term in $\tilde{E}_{11}$ in (\ref{Eortho3})
is again extremely small $V_2/|t_2|\approx 0.02$ and can be 
completely neglected.\cite{milaXZ}
Further extension of the above calculations to longer 
loops show that
tunneling matrix elements of four-dimer moves are rather small
$\sim 0.07t_1$ and should also be neglected.

\subsection{Second-order mapping}

Analysis starts again with calculation of the mean-field contribution
from $\hat{\cal H}_2$ to the ground state energy (diagonal matrix elements) 
for an arbitrary dimer state. Every 
dimer has a finite energy contribution
from $\hat{\cal H}_{2b}$: $\frac{1}{6}J^2$, while all nondimer 
bonds receive mean-field contributions from $\hat{\cal H}_{2a}$:
$-\frac{1}{12}J^2$. The total mean-field energy does not, therefore,
depend on a chosen dimer covering of a triangular lattice
and is equal to
\begin{equation}
E_{\rm MF} = \Bigl( - \frac{3}{4} - \frac{3}{8}J - \frac{1}{8}J
\Bigr) \frac{N}{3} \ ,
\label{MF}
\end{equation}
where $N$ is number of sites on a kagom\'e lattice.
The mean-field energy (\ref{MF}) is subtracted in the following from 
$\hat{\cal H}_1+\hat{\cal H}_2$, such that all diagonal matrix elements
vanish.

The off-diagonal matrix element of $\hat{\cal H}_2$
for the shortest loop $C_1$ has nonzero contributions
from  $\hat{\cal H}_{2a}$: $-\frac{1}{64}J^2$, and from
$\hat{\cal H}_{2c}$: $-\frac{1}{48}J^2$. Combining them with 
the first-order result Eq.~(\ref{c1}) we obtain
\begin{equation}
t_1 =-E_{12}=-\frac{3}{2^6}J \Bigl(1-\frac{7}{9}J\Bigr)\ .
\label{t1}
\end{equation}
Similar calculation for the loops $C_2$ and $C_3$ yields
\begin{equation}
E_{23}=\frac{3}{2^8}J\Bigl(1+\frac{1}{9}J\Bigr)\ .
\label{c23}
\end{equation}
Taking into account Eq.~(\ref{Eortho}) the tunneling matrix element
between orthogonal dimer states along the loop $C_2$ ($C_3)$ becomes
\begin{equation}
t_2 =t_3 = -\tilde{E}_{23}=-\frac{15}{2^{10}}J 
\Bigl(1-\frac{1}{15}J\Bigr)\ .
\label{t2}
\end{equation}
Loop $C_4$ also acquires a finite tunneling rate 
between the two dimer 
states in the second-order. The corresponding matrix element
is, however, small $E_{12} \approx 0.003J^2$.

\section{Discussion}

Signs of the tunneling matrix elements calculated
in the previous section have certain arbitrariness. 
\cite{rokhsar,moessner01} 
The negative sign of the resonance matrix element for 
the shortest loop $C_1$ can be changed to positive by a gauge
transformation. For this, real singlet wave-functions   
have to be multiplied by complex factors
$i^{n_r + n_{l,e} -n_{l,o}}$, where for a given dimer configuration 
$n_r$ counts the number of dimers  on links pointing
upwards and right, $n_{l,e}$ ($n_{l,o}$) counts the number
of dimers on links pointing upwards and left from sites with 
even (odd) vertical coordinates. Dimers on strictly horizontal
bonds do not contribute to the phase factor.
By this operation resonance moves along every $C_1$ loop
pick up an extra $(-1)$ factor, changing $t_1\rightarrow -t_1$.
At the same time, amplitudes of dimer tunneling along $C_2$ and $C_3$ loops
do not change sign by the above gauge transformation:
$t_{2,3}\rightarrow t_{2,3}$.
An effective QD Hamiltonian for the trimerized kagom\'e 
antiferromagnet is, therefore, dominated by the kinetic energy terms 
for resonance moves
between orthogonal dimer configurations
$|\varphi_{c_n}\rangle$ and $|\varphi'_{c_n}\rangle$
for every loop $C_n$ of three different 
types $n=1$--3 on an effective triangular
lattice:
\begin{eqnarray}
&& \hat{\cal H}_{\rm QD} = \sum_{l_n}  -t_n |\varphi_{c_n}\rangle
\langle\varphi'_{c_n}| \ , 
\label{QDTri}
\\ &&
t_1 = \frac{3}{64}J\Bigl(1-\frac{7}{9}J\Bigr), \ \ 
t_2=t_3=-\frac{15}{1024}J\Bigl(1-\frac{1}{15}J\Bigr).
\nonumber 
\end{eqnarray}
Amplitudes for 
three-dimer tunneling processes 
have no significant
smallness compared to the strongest resonance move:
$t_{2,3}/t_1\approx -0.31$ for $J\ll 1$.
The kinetic coefficients $t_n$ are differently renormalized
by the second-order processes such that importance
of three-dimer moves is further increased towards the isotropic limit:
$t_{2,3}/t_1\approx -0.5$ for $J=0.5$ and $t_{2,3}/t_1\approx -1$ as 
$J\rightarrow 1$.

The QDM (\ref{QDTri}) with only two-dimer resonances
has been studied via mapping to a frustrated 
Ising model in transverse  field.\cite{moessner01,moessner02} The ground
state of this dimer model is believed to be a crystalline 
$\sqrt{12}\times\sqrt{12}$ 
state, which consists of locally resonating dimer pairs and breaks
translational symmetry of the lattice.
Such a state should have a fully gapped excitation spectrum.
Properties of the QDM (\ref{QDTri}) with several competing dimer 
resonances have not been studied so far.
Dimer resonances along longer loops, though not very small, 
frustrate each other. The ground state of the realistic model
(\ref{QDTri}) should not be, therefore, very far from the idealized 
model with $t_1$ terms only. In particular, we expect that 
the ground state breaks certain lattice  
symmetries. The excitation spectrum is also
expected to be gapped unless a fine tuning of $t_n$ drives the system
towards an Ising-type transition point between two crystalline states.

In conclusion, the presented derivation of the QDM for a realistic 
Heisenberg spin model on trimerized kagom\'e lattice illustrates a generation
of small energy scales in frustrated quantum magnets.
The dimer resonance matrix elements in (\ref{QDTri}) are
given by small fractions of a weaker exchange constant, {\it e.g.},
$t_1\approx 0.047J$. In a wide temperature
interval $t_1\ll T\ll J$ the quantum spin systems
is described by an RVB liquid of singlet pairs.
At very low temperatures $T<t_1$ a valence bond crystal 
probably replaces an RVB state. The gap between the ground state
and the first excited singlet levels is a fraction of $t_1$.
It is extremely difficult to resolve such a tiny energy scale
in the exact numerical diagonalization of small clusters.
Similarly, 
the low-temperature
regime $T\lesssim t_1$ might be beyond experimental reach
for possible realizations
of the spin-1/2 trimerized kagom\'e antiferromagnet.
The dimer crystallization at $T=0$ is driven by local
resonances. Therefore, variational mean-field type approaches
\cite{mila,auerbach} are not capable to describe the precise
nature of the corresponding ground states.

\acknowledgments

I thank to D. A. Ivanov and G. Jackeli for valuable
discussions. The hospitality of
the Condensed Matter Theory Institute of Brookhaven 
National Laboratory during the course
of present work is gratefully acknowledged.


\begin{thebibliography}{99}

\bibitem{anderson}
P. Fazekas and P. W. Anderson, 
Philos. Mag. {\bf 30}, 423 (1974);
P. W. Anderson, Science {\bf 235}, 1196 (1987).

\bibitem{sutherland}
B. Sutherland,
Phys. Rev. B {\bf 37}, R3786 (1988).

\bibitem{rokhsar}
D. S. Rokhsar and S. A. Kivelson,
Phys. Rev. Lett. {\bf 61}, 2376 (1988).

\bibitem{moessner00}
R. Moessner, S. L. Sondhi, and P. Chandra,
Phys. Rev. Lett. {\bf 84}, 4457 (2000).

\bibitem{moessner01}
R. Moessner and S. L. Sondhi,
Phys. Rev. Lett. {\bf 86}, 1881 (2001).

\bibitem{moessner02}
R. Moessner and S. L. Sondhi,
Prog. Theor. Phys. Suppl.  {\bf 145}, 37 (2002).

\bibitem{ioselevich}
A. Ioselevich, D. A. Ivanov, and M. V. Feigelman,
Phys. Rev. B {\bf 66}, 174405 (2002).

\bibitem{book}
A. Auerbach, {\it Interacting Electrons and Quantum Magnetism}
(Springer, New York, 1994).

\bibitem{lecheminant}
P. Lecheminant, B. Bernu, C. Lhuillier, L. Pierre,
and P. Sindzingre, 
Phys. Rev. B {\bf 56}, 2521 (1997).

\bibitem{waldtmann}
C. Waldtmann, H.-U. Everts, B. Bernu, C. Lhuillier,
P. Sindzingre, P. Lecheminant, and L. Pierre,
Eur. Phys. J. B {\bf 2}, 501 (1998).

\bibitem{subrah}
V. Subrahmanyam, 
Phys. Rev. B {\bf 52}, 1133 (1995).

\bibitem{mila}
F. Mila, 
Phys. Rev. Lett. {\bf 81}, 2356 (1998).

\bibitem{raghu}
C. Raghu, I. Rudra, S. Ramasesha, and D. Sen,
Phys. Rev. B {\bf 62}, 9484 (2000).

\bibitem{auerbach}
R. Budnik and A. Auerbach,
Phys. Rev. Lett. {\bf 93}, 187205 (2004).

\bibitem{santos}
L. Santos, M. A. Baranov, J. I. Cirac, H.-U. Everts, H. Fehrmann, 
and M. Lewenstein, 
Phys. Rev. Lett. {\bf 93}, 030601 (2004).

\bibitem{mambrini}
M. Mambrini and F. Mila,
Eur. Phys. J. B {\bf 17}, 651 (2000).

\bibitem{syromyat}
A. V. Syromyatnikov and S. V. Maleyev,
Phys. Rev. B {\bf 66}, 132408 (2002).

\bibitem{milaXZ}
similar results have recently been obtianed by
F. Mila et al., private communication (2004).

\end{thebibliography}
\end{document}